\documentclass{article}

\usepackage{arxiv}

\usepackage[utf8]{inputenc} % allow utf-8 input
\usepackage[T1]{fontenc}    % use 8-bit T1 fonts
\usepackage{hyperref}       % hyperlinks
\usepackage{url}            % simple URL typesetting
\usepackage{booktabs}       % professional-quality tables
\usepackage{amsfonts}       % blackboard math symbols
\usepackage{nicefrac}       % compact symbols for 1/2, etc.
\usepackage{microtype}      % microtypography
\usepackage{lipsum}
\usepackage{graphicx}
\usepackage{tabularx}
\usepackage{authblk}
\usepackage{amsmath} 
\usepackage{booktabs}
\usepackage{cite}
\graphicspath{ {./images/} }

\title{An Iterative LLM Framework for SIBT utilizing RAG-based Adaptive Weight Optimization}

\author[1]{Zhuo Xiao}
\author[1]{Fugen Zhou}
\author[1]{Qinglong Yao}
\author[1]{Jingjing Wang}
\author[1]{Bo Liu}
\author[2]{Haitao Sun}
\author[2]{Zhe Ji}
\author[2]{Yuliang Jiang}
\author[2]{Junjie Wang}
\author[3]{Qiuwen Wu}

\affil[1]{Image Processing Center, Beihang University, Beijing 100191, China}
\affil[2]{Department of Radiation Oncology, Peking University Third Hospital, Beijing 100191, China}
\affil[3]{Department of Radiation Oncology, Duke University Medical Center, Durham, NC 27710, USA}
\affil[ ]{\textbf{\texttt{bo.liu@buaa.edu.cn}}}

\begin{document}
\maketitle
\begin{abstract}
Seed implant brachytherapy (SIBT) is an effective cancer treatment modality; however, clinical planning often relies on manual adjustment of objective function weights, leading to inefficiencies and suboptimal results. This study proposes an adaptive weight optimization framework for SIBT planning, driven by large language models (LLMs). A locally deployed DeepSeek-R1 LLM is integrated with an automatic planning algorithm in an iterative loop. Starting with fixed weights, the LLM evaluates plan quality and recommends new weights in the next iteration. This process continues until convergence criteria are met, after which the LLM conducts a comprehensive evaluation to identify the optimal plan. A clinical knowledge base, constructed and queried via retrieval-augmented generation (RAG), enhances the model’s domain-specific reasoning. The proposed method was validated on 23 patient cases, showing that the LLM-assisted approach produces plans that are comparable to or exceeding clinically approved and fixed-weight plans, in terms of dose homogeneity for the clinical target volume (CTV) and sparing of organs at risk (OARs). The study demonstrates the potential use of LLMs in SIBT planning automation.
\end{abstract}

\section{Introduction}
Seed implant brachytherapy (SIBT) is an established and effective treatment modality for tumors at many site\cite{cuaEvolvingLandscapeHead2024}\cite{rodinSystematicReviewTreating2018}. The core principle of SIBT involves precisely delivering a high radiation dose to the tumor while minimizing exposure to surrounding healthy tissues. This is achieved by carefully planning safe needle trajectories and optimizing the placement of radioactive seeds. To enhance precision and improve treatment outcomes, the CT-guided dynamic dose planning workflow has been developed. This workflow typically encompasses three key phases: preoperative imaging-based planning (often incorporating 3D-printed templates for guidance), intraoperative CT imaging for real-time adjustments and adaptive optimization, and postoperative imaging for comprehensive dose verification \cite{jiEffectivenessPrognosticFactors2019}\cite{jiangSideEffectsCTguided2018}\cite{wangExpertConsensusWorkshop2017}. In this study, we focus on the preoperative planning phase.

A critical step in SIBT planning, once the needle paths are determined, is the optimization of seed positions using dose-based inverse planning methods. This process is commonly formulated and solved as an optimization problem\cite{liangNovelGreedyHeuristicbased2015a}\cite{guthierNewOptimizationMethod2015}\cite{villaFastMonteCarloBased2022a}\cite{xiaoAutomaticPlanningHead2024}. Clinical objectives in SIBT are often multifaceted and can be interrelated or even conflicting. These objectives typically include achieving adequate dose coverage of the clinical target volume (CTV), adhering to strict dose constraints for organs at risk (OARs), and minimizing the number of needles and seeds used. A primary challenge in this optimization lies in effectively translating these diverse clinical objectives into a cohesive, weighted objective function that can be solved by optimization algorithms.

Currently, this challenge is addressed through trial-and-error adjustments, which are highly dependent on planner expertise and remain time-consuming.Several automated weight-tuning strategies have been proposed previously, many depend on handcrafted optimization pipelines\cite{zhangInversePlanningSimulated2022}, simplistic reward functions\cite{puDeepReinforcementLearning2022}\cite{shenIntelligentInverseTreatment2019}, or multi-objective optimization models to approximate the Pareto frontier\cite{jafarzadehPenaltyWeightTuning2024}. However, these methods lack the capacity to semantically interpret clinical objectives and offer limited adaptability to dynamic and patient-specific planning scenarios.

In recent years, large language models (LLMs) and multimodal large language models (MLLMs) have been explored for weight tuning in radiotherapy treatment planning, showing promising potential \cite{wangFeasibilityStudyAutomating2025}\cite{liu2025automatedradiotherapytreatmentplanning}. For example, Liu et al. proposed GPT-RadPlan, a GPT-4V-based agent that emulates human planners by iteratively adjusting objective weights and doses based on DVH analysis and textual feedback\cite{liu2025automatedradiotherapytreatmentplanning}.  However, current approaches face limitations. Primarily, they rely on general-purpose LLMs that lack integration with structured clinical knowledge, limiting their ability to make context-aware decisions. Furthermore, most implementations depend on cloud-based APIs (e.g., ChatGPT-4\cite{openaiGPT4TechnicalReport2024}), which are unsuitable for offline or network-restricted clinical environments.

To address these limitations and further explore LLM-based automatic planning, we propose a novel framework that leverages the reasoning capabilities of LLMs to iteratively optimize weights in SIBT planning. This approach integrates a locally deployed retrieval-augmented generation (RAG) knowledge base with an LLM-driven evaluation and refinement workflow. To the best of our knowledge, this is the first study to explore the use of LLMs for treatment plan optimization in brachytherapy. The main contributions of this work are as follows:

\begin{itemize}
    \item We construct a domain-specific, locally hosted RAG knowledge base tailored for SIBT planning, enabling the LLM to dynamically incorporate clinical knowledge during plan evaluation and weight recommendation, while ensuring data privacy and offline operability.
    
    \item We develop a fully local, LLM-guided iterative workflow that integrates plan evaluation and automated weight adjustment. The standardized output schema allows seamless interaction between the language model and the optimization solver.
    
    \item We demonstrated that this method outperforms fixed-weight baselines and achieves planning quality comparable to clinical plans, with improved sparing of OAR and higher efficiency.

\end{itemize}

\section{Related Works}
\subsection{Weight Tuning of Treatment Planning}
Traditionally, weight tuning in SIBT has been carried out using protocol-based methods\cite{zhangInversePlanningSimulated2022}. This approach simulates the iterative decision-making process of human planners by starting with predefined clinical objectives and iteratively adjusting these objectives in a flowchart-like manner based on plan quality feedback. In the context of high-dose-rate (HDR) brachytherapy, some researchers have formulated weight tuning as a multi-objective optimization problem, applying Bayesian optimization to identify Pareto-optimal combinations of penalty weights\cite{jafarzadehPenaltyWeightTuning2024}. Other studies have explored the use of reinforcement learning (RL) for automated weight tuning\cite{puDeepReinforcementLearning2022}\cite{shenIntelligentInverseTreatment2019}; however, the design of appropriate reward functions and action spaces remains a significant challenge, limiting the practical effectiveness of RL-based approaches. In the domain of external beam radiotherapy, LLMs have demonstrated the ability to emulate planner behavior in adjusting optimization weights with a limited number of clinical planning examples, including text, DVH tables, and dose distribution images\cite{wangFeasibilityStudyAutomating2025}\cite{liu2025automatedradiotherapytreatmentplanning}. Despite promising early results, such approaches remain largely unexplored in brachytherapy and face challenges in generalizing across diverse cases due to limited domain knowledge integration. Moreover, most existing LLM-based planning methods rely on cloud-hosted APIs, which raises concerns regarding data security, network dependency, and regulatory compliance, thereby limiting their suitability for clinical deployment.

\subsection{Large Language Models}
Recent advances in large language models, such as DeepSeek-R1\cite{deepseek-aiDeepSeekR1IncentivizingReasoning2025}, Gemini\cite{teamGeminiFamilyHighly2025}, GPT-4\cite{openaiGPT4TechnicalReport2024}, LLaMA\cite{grattafioriLlama3Herd2024}, and QWen\cite{yangQwen3TechnicalReport2025}, have demonstrated remarkable capabilities in instruction following\cite{zhouInstructionFollowingEvaluationLarge2023}, contextual learning\cite{rouhiEnhancingObjectDetection2025}, and chain-of-thought reasoning\cite{weiChainofThoughtPromptingElicits}. Trained on massive general-domain corpora, these models exhibit emergent behaviors such as in-context learning and few-shot generalization, allowing for integration into downstream workflows with minimal task-specific supervision. However, due to their reliance on publicly available internet data, general-purpose LLMs often underperform in specialized domains such as healthcare, finance, or law, where domain-specific terminology and reasoning are critical.

To address this, domain-adapted LLMs have been developed through pretraining or fine-tuning on specialized datasets. For instance, in the medical field, models such as BioGPT\cite{luoBioGPTGenerativePretrained2022}, Med-PaLM\cite{singhalExpertlevelMedicalQuestion2025}, and LLaVA-Med\cite{liLLaVAMedTrainingLarge} integrate structured biomedical knowledge and clinical corpora to improve performance on tasks such as clinical question answering\cite{singhalExpertlevelMedicalQuestion2025}, radiology report generation\cite{sloanAutomatedRadiologyReport2025}, and image-guided diagnosis\cite{liuMedicalMultimodalLarge2023}. While effective, these approaches demand significant computational resources, large volumes of annotated data, and frequent retraining to remain current. This limits their scalability and flexibility in practical clinical settings.

To overcome these limitations and maintain adaptability without incurring high retraining costs, our proposed approach utilizes a general-purpose LLM that dynamically accesses external domain knowledge through retrieval-augmented generation (RAG). Instead of embedding domain knowledge into model parameters, RAG retrieves relevant external information during inference to supplement generation, offering improved transparency and up-to-date knowledge access. This hybrid approach has shown promise in medical question answering and decision support\cite{liuMedicalMultimodalLarge2023}\cite{zhangLeveragingLongContext2025}\cite{alonsoMedExpQAMultilingualBenchmarking2024}\cite{wangRetCareInterpretableClinical}\cite{sRAGbasedMedicalAssistant2024}, and forms the foundation of our strategy for enabling general-purpose LLMs to reason over clinical data without domain-specific fine-tuning.

\section{Methods}
\subsection{Framework Overview}
\begin{figure*}[htbp]
    \centering
    \includegraphics[width=\textwidth]{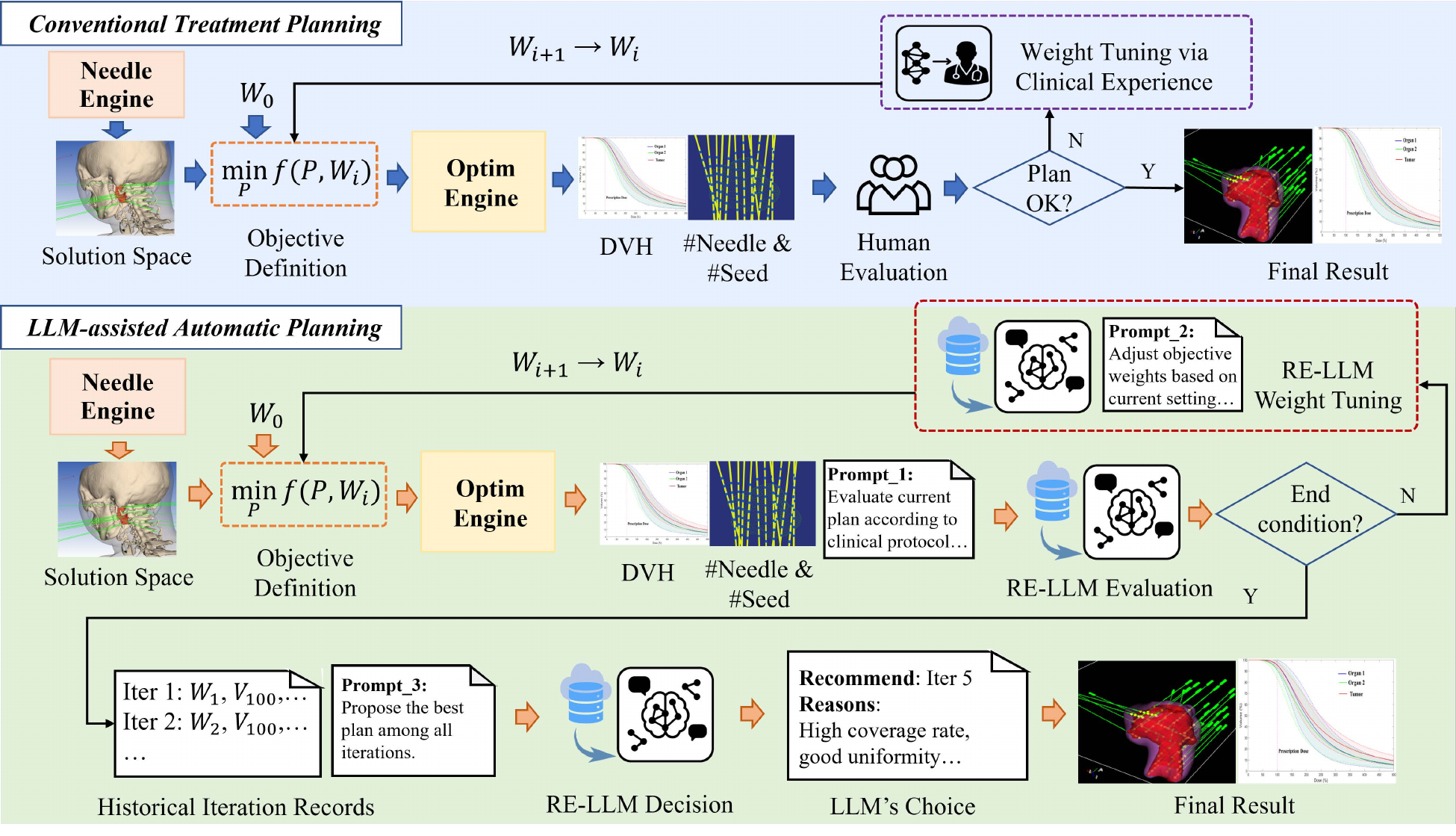} 
    \caption{The overall framework of the proposed method.} 
    \label{fig:fig1} 
\end{figure*}

SIBT planning procedure typically involves several key steps. First, based on the patient’s anatomical structure, a set of candidate needle trajectories is generated either manually by physicians or through automated methods. Since radioactive seeds are implanted along the needles, this step effectively defines the solution space. Subsequently, an objective function is formulated, usually incorporating weighted conflicting terms for CTV coverage, OAR sparing, and resource cost (e.g., number of needles and seeds). An optimization engine is then employed to solve this objective, producing a candidate treatment plan. In the conventional paradigm, as shown in Fig.~\ref{fig:fig1}, the planning result is evaluated by the physician and the weights are manually adjusted based on physician experience. This iterative weight tuning is repeated until a satisfactory plan is achieved, a process that is time-consuming and subject to inter-operator variability.

In contrast to conventional workflows, the proposed framework employs a retrieval-augmented large language model (RE-LLM) to automate both optimization weight tuning and treatment plan evaluation. By integrating a RAG mechanism, the RE-LLM dynamically accesses locally stored clinical guidelines and physician-defined planning rules to support protocol-compliant, informed decision-making. Starting from an initial set of weights, the model iteratively evaluates the generated plans and suggests weight modifications to address identified suboptimalities. This process mimics the reasoning strategy of experienced planners and allows the system to adaptively refine candidate plans in a closed-loop. Detailed strategies for weight adjustment and multi-plan selection are described in the following sections.

\subsection{LLM-based Optimization}
In the following, we describe three essential components of the proposed LLM-based optimization framework, which are plan evaluation, weight tuning and multi-plan comparison and selection.
\subsubsection{Plan Evaluation}
Once the optimization engine generates a treatment plan using a given set of objective weights, relevant parameters are extracted and stored. These include the DVH metrics for the CTV and OARs, as well as the number of implanted seeds and needles. Historical optimization records are also preserved and incorporated into the plan evaluation prompt submitted to the LLM.

To support accurate and context-aware reasoning, the LLM receives structured input that contains tabular representations of DVH data, key dosimetric values, and implant-related information. This approach avoids the limitations of current multimodal LLMs, which have difficulty interpreting DVH images directly \cite{wangFeasibilityStudyAutomating2025}. Presenting the data in a text-based tabular format allows the model to interpret quantitative information more reliably. In addition, clinical thresholds and decision rules defined by physicians are stored in a local structured knowledge base. These are retrieved using a retrieval-augmented generation mechanism to inform the LLM during evaluation. This integration ensures that the model’s assessment aligns with established clinical standards and decision-making criteria.

During evaluation, the LLM analyzes the provided dosimetric inputs to identify key deficiencies, such as suboptimal CTV coverage or excessive OAR dose, based on both absolute thresholds and relative trade-offs. The model also compares the current plan against historical optimization results to recognize performance trends and detect repeated failure patterns. By integrating these multiple sources of information, the LLM performs a qualitative assessment of plan quality and generates natural language feedback to guide subsequent weight adjustments. This reasoning process allows the system to iteratively refine treatment planning in alignment with clinical intent.

The closed-loop process continues until one of the following conditions is met: (1) CTV coverage satisfies clinical criteria, but further tuning fails to reduce OAR dose without deteriorating the CTV coverage; (2) the needle count cannot be further reduced without deteriorating the dose distribution for both OARs and CTV; (3) the maximum number of iterations is reached, which is set to 10 in this study.

\subsubsection{Weight Tuning}
If the termination condition is not met, the system proceeds to adjust the objective function weights based on the LLM's evaluation of the current plan. The tuning process is guided by configurable clinical preferences that prioritize adequate CTV coverage to ensure therapeutic efficacy, while also considering OAR sparing and minimizing procedural complexity, such as the number of implanted needles.

While the exact weight adjustment is dynamically determined by the LLM, general clinical preferences are also incorporated to inform its strategy through RAG mechanism. For example, if a critical metric such as CTV coverage is far from acceptable, the corresponding weight can be significantly increased. For secondary metrics, such as needle reduction, moderate upward adjustments can be applied when the primary objectives are already satisfied, allowing for further optimization without compromising overall plan quality. In cases where two consecutive iterations yield no observable improvement, the adjustment range may be expanded to help the optimizer escape local optima.If no improvement is observed after three consecutive iterations, the process is considered to have met the stopping criterion.

\subsubsection{Multi-Plan Comparison and Selection}
After the tuning loop concludes, all candidate plans generated during the process are reviewed and compared by the LLM. Rather than relying solely on the outcome of the final iteration, the system performs a holistic, knowledge-guided evaluation of all intermediate results. This step mitigates the uncertainty inherent in heuristic optimization, where not every weight adjustment guarantees improvement.

The multi-plan evaluation considers quantitative dosimetric metrics and clinical priorities retrieved from the local knowledge base. Each plan is assessed across multiple dimensions, including CTV coverage and dose uniformity, OAR sparing, and the efficiency of needle and seed utilization. Based on these multi-criteria assessments, the LLM selects the plan that best reflects the clinical priority hierarchy and trade-offs. This final selection ensures consistency, robustness, and clinical interpretability across patients. It also enables scalable and automated treatment planning guided by expert-level decision logic.

\subsection{RAG-enhanced Prompt Generation}
Prompt engineering plays a pivotal role in guiding LLMs to perform reliably across complex, domain-specific tasks such as clinical treatment planning\cite{marvinPromptEngineeringLarge2024}. Given the inherent ambiguity and variability in natural language, a well-designed prompt is essential to constrain the model’s reasoning, reduce inconsistency, and ensure alignment with clinical expectations. The template follows a rigid and standardized format, explicitly presenting all relevant clinical input data and guiding the model through evaluation and decision-making steps. This design improves the reproducibility of model outputs, enhances compatibility with clinical protocols, and ensures stable performance across diverse patient cases.

\begin{figure}[htbp]
    \centering
    \includegraphics[width=0.8\linewidth]{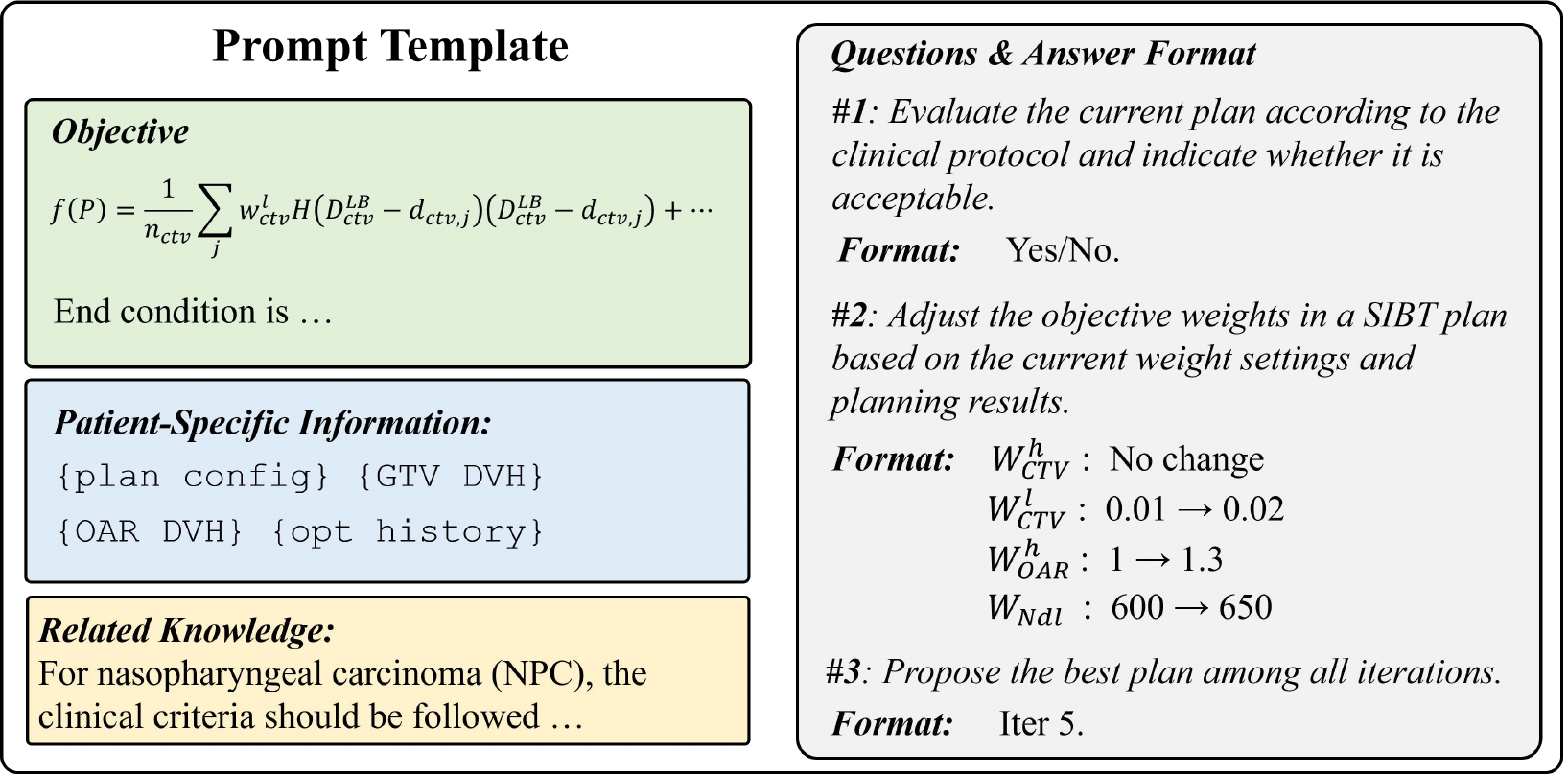} 
    \caption{Prompt template for weight optimization. It includes task objectives, patient-specific inputs (e.g., DVH and plan history), related clinical knowledge, and a standardized Q\&A format for weight adjustment, plan evaluation, and selection.} 
    \label{fig:fig2} 
\end{figure}

As shown in Fig.~\ref{fig:fig2}, the left side of the template provides contextual input, including the objective function structure, patient-specific planning data and relevant clinical knowledge retrieved via the RAG module. This integrated context ensures that the model's reasoning remains aligned with protocol-defined criteria and patient-specific conditions. The right side of the template defines a standardized question-and-answer format, which instructs the model to perform weight adjustment, assess plan acceptability, and select the optimal plan among all iterations. Each response adheres to a fixed schema designed for seamless parsing by the downstream optimization engine.

\begin{figure}[htbp]
    \centering
    \includegraphics[width=0.8\linewidth]{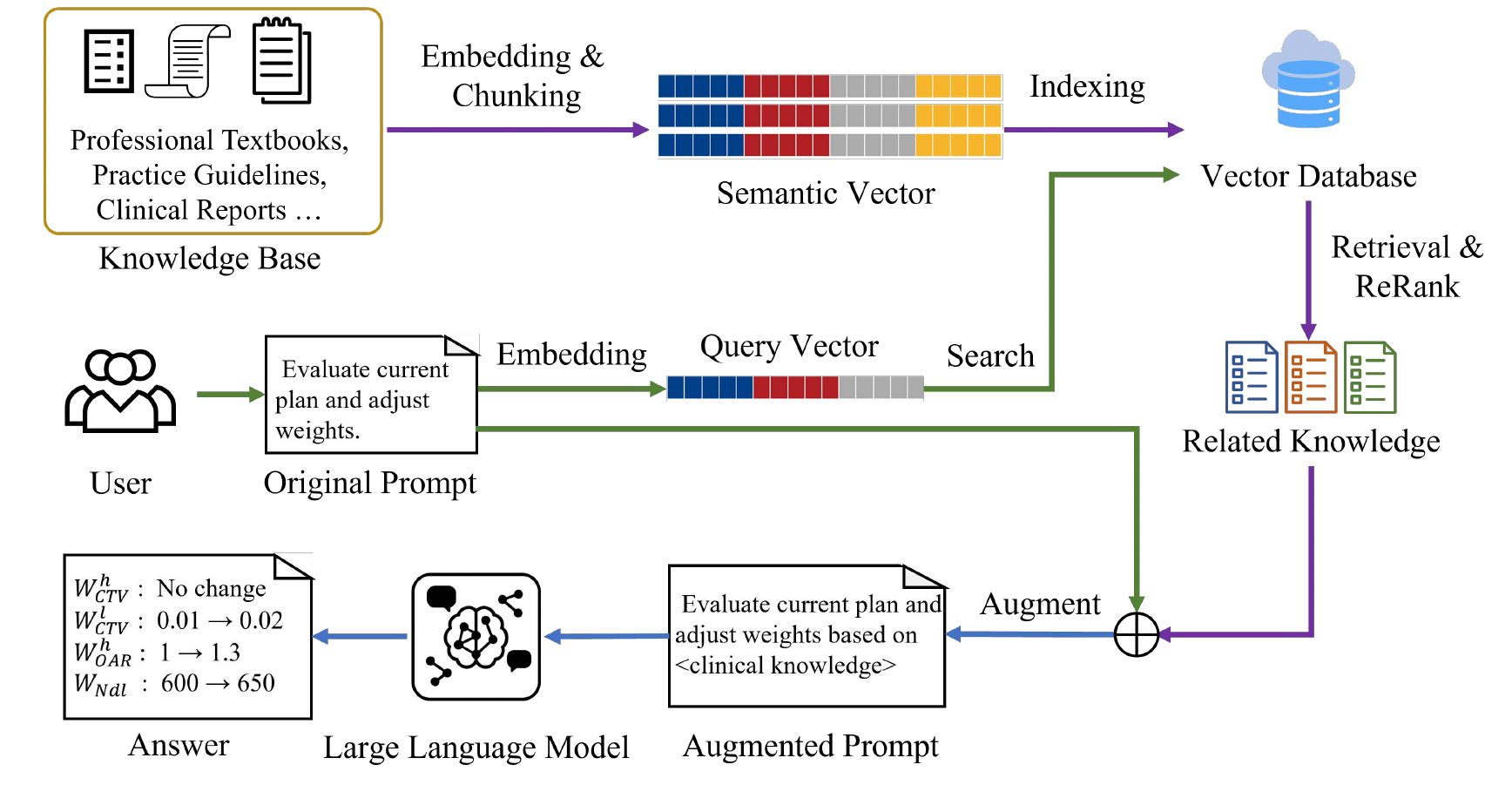} 
    \caption{The framework of RAG-enhanced prompt generation.} 
    \label{fig:fig3} 
\end{figure}

To compensate for the domain knowledge limitations of the base LLM, a RAG module was incorporated into the proposed framework to dynamically inject SIBT-specific clinical knowledge during prompt generation and decision making. As illustrated in Fig.~\ref{fig:fig3}, the RAG system consists of two key components: a structured clinical knowledge base and a hybrid retrieval pipeline. During each iteration, the system queries relevant protocol and case-specific knowledge, which is then incorporated into the prompt to guide the LLM in evaluating the current plan and recommending clinically informed weight adjustments.

The knowledge base was constructed from professional textbooks, practice guidelines, and clinical reports relevant to SIBT. Source documents in PDF and plain text formats were parsed and segmented using a two-stage splitting strategy. First, an embedding-guided semantic chunking approach preserved contextual coherence; subsequently, content-aware secondary splitting optimized chunk size for both retrieval accuracy and generative performance.

For knowledge retrieval, a hybrid retriever combining dense and sparse retrieval techniques was adopted. Dense retrieval utilized semantic embeddings generated by the BGE-Small-Zh-v1.5 model\cite{bge_m3} and stored in a Chroma-powered vector database for efficient similarity search. Given a query $q$, the dense similarity score is calculated via cosine similarity between the query embedding $\mathbf{v}_q$ and document embeddings $\mathbf{v}_i$:

\begin{equation}
S_{\text{d}}(q, i) = \frac{\mathbf{v}_q \cdot \mathbf{v}_i}{\|\mathbf{v}_q\| \|\mathbf{v}_i\|}
\end{equation}

For sparse retrieval, the BM25 algorithm\cite{robertsonProbabilisticRelevanceFramework2009} was utilized to complement the dense retrieval results. The BM25 score is computed as:

\begin{equation}
S_{\text{b}}(q, D) = \sum_{t \in q} \! \text{IDF}(t)\!\cdot\!
\frac{f(t,D)\cdot(k_1+1)}{f(t,D) + k_1 \left(1 - b + b \cdot \frac{|D|}{\text{avgdl}}\right)}
\end{equation}

where $t$ denotes a term in the query $q$, and $f(t, D)$ is the frequency of term $t$ in document $D$. The parameter $|D|$ indicates the length of document $D$, while $\text{avgdl}$ denotes the average document length across the entire corpus. The term $\text{IDF}(t)$ reflects the inverse document frequency of term $t$, which down-weights common terms and emphasizes more informative ones. The hyperparameter $k_1$ controls the scaling of term frequency contribution, while $b$ adjusts for document length normalization. In our experiments, $k_1 = 1.2$, $b = 0.75$.

The final retrieval score was obtained by weighted fusion of dense and sparse scores. To further refine retrieval precision, a cross-encoder reranking model \cite{nogueiraPassageRerankingBERT2020} was applied to re-score candidate documents based on query-document relevance pairs. Only documents with final scores exceeding a predefined threshold are selected for retrieval and passed to the language model for subsequent reasoning.

During the weight adjustment process in each iteration, the RAG module queries and retrieves knowledge specific to the tumor site and OARs involved in the current case, enabling clinically informed evaluations and adaptive weight recommendations.

\section{Experiments}
\subsection{Patient data and preprocessing}
A retrospective cohort of plan data was collected from 23 H\&N patients who had undergone SIBT. The study was approved by the Institutional Ethics Committee of Peking University Third Hospital (Beijing, China; Approval No.M2021438). Each patient's data included their planning CT, contours of the CTV and OARs, and comprehensive treatment plan information such as needle paths, seed positions, air-kerma strength, and dose prescription.

Among the 23 cases, primary tumor locations varied. Seven patients (30\%) presented with parotid tumors, four (17\%) with oral or oropharyngeal tumors, four (17\%) with nasopharyngeal tumors, three (13\%) with primary cervical neoplasms, two (9\%) with orbital or periorbital tumors, one (4\%) with laryngeal or hypopharyngeal tumors, one (4\%) with a nasal or paranasal sinus tumor, and one (4\%) with an infratemporal fossa tumor. The number and type of OARs varied according to tumor location. A total of eight patients had their OARs contoured, which included the trachea, esophagus, spinal cord, optic nerve, eyeball, lens, parotid gland, subclavian artery, internal carotid artery, and other critical vascular structures. The prescription dose of the plans ranges between 80 and 120 Gy. The number of needles in each plan ranges from 4 to 20 and seeds from 14 to 58. All plans use I-125 seeds of air-kerma strength in the range of 0.381-0.889 U and each plan only contains seeds of the same strength. The original CT images have varying resolutions (in-slice resolution: 0.47 mm to 1.17 mm; slice thickness: 1.0 mm to 5.0 mm).

\subsection{Evaluation and Implementation}
To quantitatively evaluate the performance of the proposed workflow, we compared planning results from three approaches: (1) clinical plans, which involved manual needle path design and manual adjustment of objective function weights; (2) fixed-weight optimization, which incorporated automated needle path design with fixed target weights; and (3) the proposed LLM-enabled adaptive-weight optimization, which built upon the fixed-weight plans by applying the proposed adaptive weight adjustment strategy using a large language model. For the CTV, we statistically evaluated dose indices including $D_{90}$, $V_{100}$, $V_{150}$, and $V_{200}$. For OARs, evaluation was performed using parameters including $V_{50}$, $D_{\text{mean}}$, $D_{\text{max}}$, $D_1$, $D_{1\text{cc}}$, and $D_{0.1\text{cc}}$.

To define the solution space, needle trajectories were generated following the approach described in our previous work \cite{xiaoAutomaticPlanningHead2024}. For objective definition, a penalty-driven cost function was employed, defined as:

\begin{align}
f(P) &= \sum_{o \in \{\text{ctv}, \text{oar}\}} \frac{1}{n_o} \sum_j w_o^h \, H(d_{o,j} - D_o^{UB}) (d_{o,j} - D_o^{UB}) \notag \\
&\quad + \frac{1}{n_{\text{ctv}}} \sum_j w_{\text{ctv}}^l \, H(D_{\text{ctv}}^{LB} - d_{\text{ctv},j}) (D_{\text{ctv}}^{LB} - d_{\text{ctv},j}) \notag \\
&\quad + w_n N(P)
\end{align}

The function penalizes CTV underdose, CTV/OAR overdose, and excessive needle usage. The first term applies to dose violations in both CTV and OAR structures;the second addresses underdosage by penalizing voxels receiving doses below the lower bound; and the third penalizes the total number of implanted needles to encourage minimally invasive plans. The Heaviside function $H(\cdot)$ activate penalties upon constraint violation. For the CTV, the lower and upper bounds are set to 1 and 2 times the prescription dose, respectively; for each OAR, the upper bound is set to 1 times the prescription dose. \( P \) represents the seed and needle configuration. Initial fixed weights \( w_{\text{ctv}}^l \), \( w_{\text{ctv}}^h \), \( w_{\text{oar}}^h \) and \( w_n \) were set to 20, 0.01, 1 and 600. The number of tunable weights ranges from 3 to 6, depending on the number of OARs involved.

The optimization engine uses a heuristic strategy involving iterative seed addition and removal. In each iteration, the algorithm temporarily adds a seed to every unoccupied candidate position, evaluates the objective function, and permanently retains the seed yielding the largest cost reduction. This is followed by a removal phase, where each existing seed is temporarily deleted, and the one whose removal produces the greatest cost reduction is permanently excluded. The process repeats until no further improvement is achieved, at which point the optimization terminates.

The core algorithms for needle trajectory generation and optimization were developed using C/C++ and CUDA 11.8. For dose calculations, we adhered to the AAPM TG-43 formalism\cite{nathDosimetryInterstitialBrachytherapy1995}\cite{rivardUpdateAAPMTask2004a} to ensure consistency with the clinical reference plans, allowing for a fair and direct comparison. The dose computation was substantially accelerated through CUDA-based GPU parallel processing.

For the implementation of RAG, text data and documents were processed and loaded using the LangChain library\cite{Chase_LangChain_2022}.We used Python's pandas library\cite{mckinneyDataStructuresStatistical2010} to parse CSV files of DVH, extracting data and populating prompt placeholders to construct the LLM's input. For LLM deployment, the DeepSeek-R1-Distill-Qwen-14B\cite{deepseekai2025deepseekr1incentivizingreasoningcapability} model was deployed on an NVIDIA A100 (80 GB VRAM) GPU for inference, leveraging the Hugging Face Transformers library\cite{wolfTransformersStateoftheArtNatural2020}. Model inference was performed using half-precision (float16) to improve computational efficiency.

\subsection{Comparison Experiments}
\begin{figure}[htbp]
    \centering
    \includegraphics[width=0.8\linewidth]{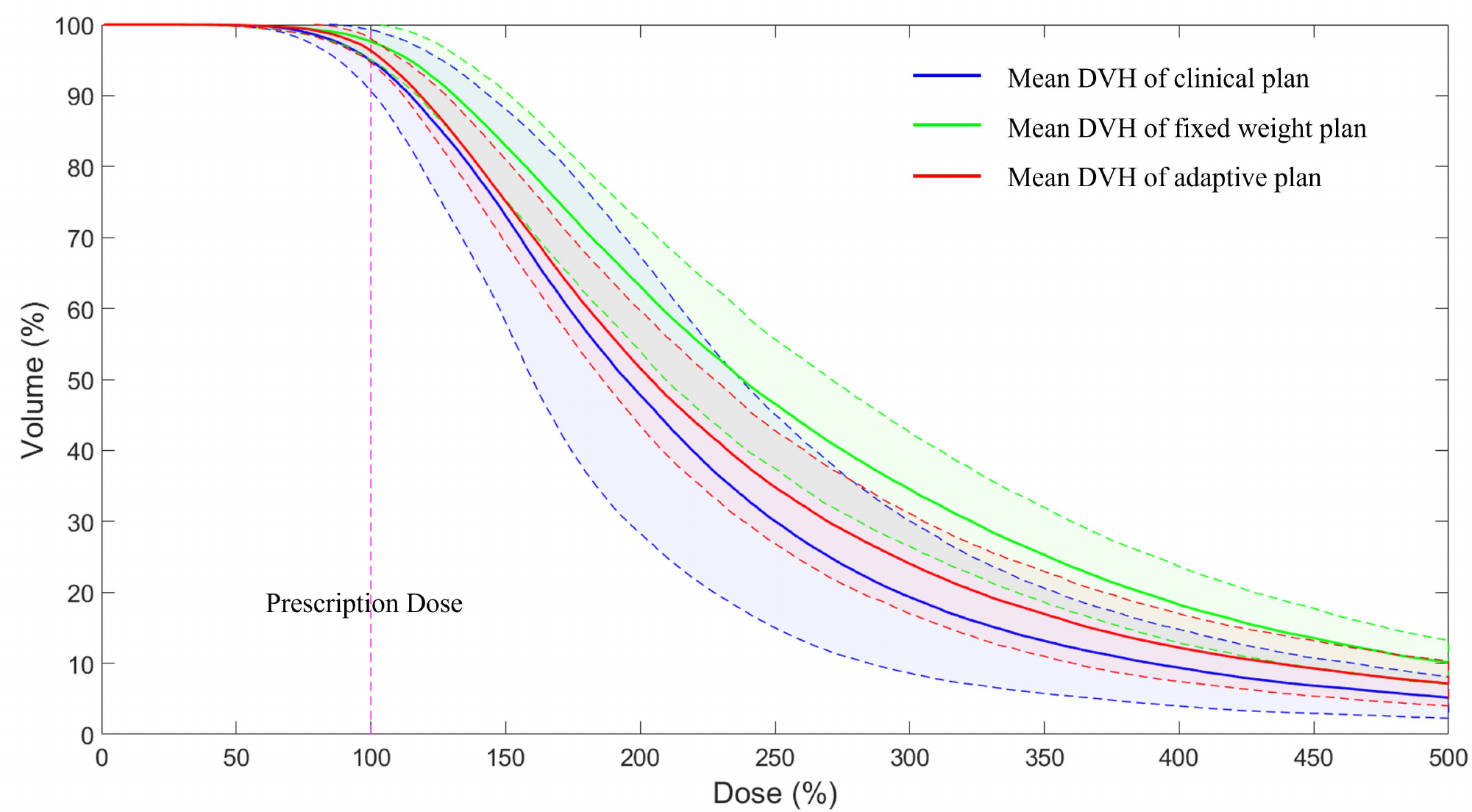} 
    \caption{Comparison of mean CTV DVH and DVH SD of all patient cases.} 
    \label{fig:fig4} 
\end{figure}

Table~\ref{tab:CTV_comparison} summarizes the quantitative results for different planning strategies. Fig.~\ref{fig:fig4} shows the mean DVHs and corresponding standard deviation bands for the CTV across the three approaches. Compared to clinical plans, the proposed method achieved comparable DVH metrics while requiring fewer needles. Furthermore, the adaptive plans demonstrated reduced variability, suggesting improved planning stability. Compared with fixed-weight planning, the proposed approach yielded better dose homogeneity in the CTV, with $V_{150}$ and $V_{200}$ not being excessively high.

\begin{table*}[htbp]
\centering
\caption{Comparison of CTV dosimetric parameters and resource usage among clinical plans, fixed-weight inverse plans, and adaptive plans.}
\begin{tabular}{lccccc}
\toprule
Plans & $V_{100}$ (\%)↑ & $V_{150}$ (\%)↓ & $V_{200}$ (\%)↓ & $D_{90}$ (\%)↑ & \#needle ↓ \\
\midrule
Clinical plans & $94.9 \pm 4.3$ & $73.0 \pm 15.1$ & $47.8 \pm 19.5$ & $119.5 \pm 17.3$ & $10 \pm 3$ \\
Fixed-weight plans & $97.6 \pm 2.6$ & $82.9 \pm 7.6$ & $63.1 \pm 9.1$ & $133.1 \pm 12.8$ & $9 \pm 2$ \\
Adaptive plans & \textbf{96.4 $\pm$ 1.6} & \textbf{75.2 $\pm$ 6.2} & \textbf{51.8 $\pm$ 8.6} & \textbf{120 $\pm$ 7.9} & \textbf{9 $\pm$ 2} \\
\midrule
$p$ value (CP vs AP) & $0.2$ & $0.56$ & $0.41$ & $0.9$ & $0.02$ \\
$p$ value (FWP vs AP) & $0.23$ & $<0.001$ & $<0.001$ & $<0.001$ & $0.27$ \\
\bottomrule
\end{tabular}

\vspace{1ex}
\begin{flushleft}
\textit{Note}: $P$ values are from two-sided Wilcoxon signed-rank tests comparing adaptive plans with clinical plans (CP vs AP) and fixed-weight plans (FWP vs AP).
\end{flushleft}
\label{tab:CTV_comparison}
\end{table*}

\begin{table*}[htbp]
\centering
\caption{Comparison of OAR dosimetric parameters among clinical plans, fixed-weight inverse plans, and adaptive plans.}
\begin{tabular}{lcccccc}
\toprule
Plans & $V_{50}$ (\%)↓ & $D_{\text{mean}}$ (Gy)↓ & $D_{\text{max}}$ (Gy)↓ & $D_1$ (Gy)↓ & $D_{1\text{cc}}$ (Gy)↓ & $D_{0.1\text{cc}}$ (Gy)↓ \\
\midrule
Clinical plans & $41.7 \pm 20.9$ & $52.5 \pm 19.2$ & $140.7 \pm 57.1$ & $129.9 \pm 50.5$ & $52.8 \pm 22.1$ & $99.7 \pm 25.9$ \\
Fixed-weight plans & $42.0 \pm 21.3$ & $48.7 \pm 17.4$ & $116.6 \pm 38.8$ & $106.9 \pm 33.4$ & $50.8 \pm 24.6$ & $89.3 \pm 29.2$ \\
Adaptive plans & \textbf{36.3 $\pm$ 22.2} & \textbf{44.8 $\pm$ 16.5} & \textbf{109.5 $\pm$ 36.9} & \textbf{98.5 $\pm$ 29.8} & \textbf{47.0 $\pm$ 22.2} & \textbf{81.6 $\pm$ 25.3} \\
\midrule
$p$ value (CP vs AP) & $0.25$ & $0.05$ & $0.11$ & $0.02$ & $0.2$ & $0.05$ \\
$p$ value (FWP vs AP) & $0.05$ & $0.15$ & $0.74$ & $0.11$ & $0.25$ & $0.2$ \\
\bottomrule
\end{tabular}
\vspace{1ex}
\begin{flushleft}
\textit{Note}: $P$ values are from two-sided Wilcoxon signed-rank tests comparing adaptive plans with clinical plans (CP vs AP) and fixed-weight plans (FWP vs AP).
\end{flushleft}
\label{tab:OAR_comparison}
\end{table*}

Table~\ref{tab:OAR_comparison} summarizes the quantitative analysis results for OARs across the clinical plans, fixed-weight plans, and adaptive plans. Compared to the fixed-weight approach, the adaptive plans consistently achieved lower OAR doses across most metrics, indicating improved of OAR sparing. In particular, reductions in high-dose exposure metrics (e.g., $\mathrm{D}_{1}$, $\mathrm{D}_{0.1\,\mathrm{cc}}$) suggest that the adaptive strategy better controls dose hotspots within OARs. While the differences between adaptive and clinical plans are less pronounced, the adaptive approach still demonstrated comparable or improved performance in several parameters. These findings highlight the effectiveness of the LLM-guided weight tuning framework in balancing target coverage with OAR protection.

The optimization process converged within 5.3 ± 0.8 iterations (ranging from 4 to 7), with the final selected plan located at iteration 3.2 ± 1.4. Across all test cases, none reached the predefined maximum of 10, indicating that the proposed framework efficiently identifies clinically favorable solutions within a limited number of iterations. Notably, the selected plan typically emerged about two iterations before termination, implying that the final steps mainly refine marginal trade-offs and validate convergence. 

\begin{figure}[htbp]
    \centering
    \includegraphics[width=0.8\linewidth]{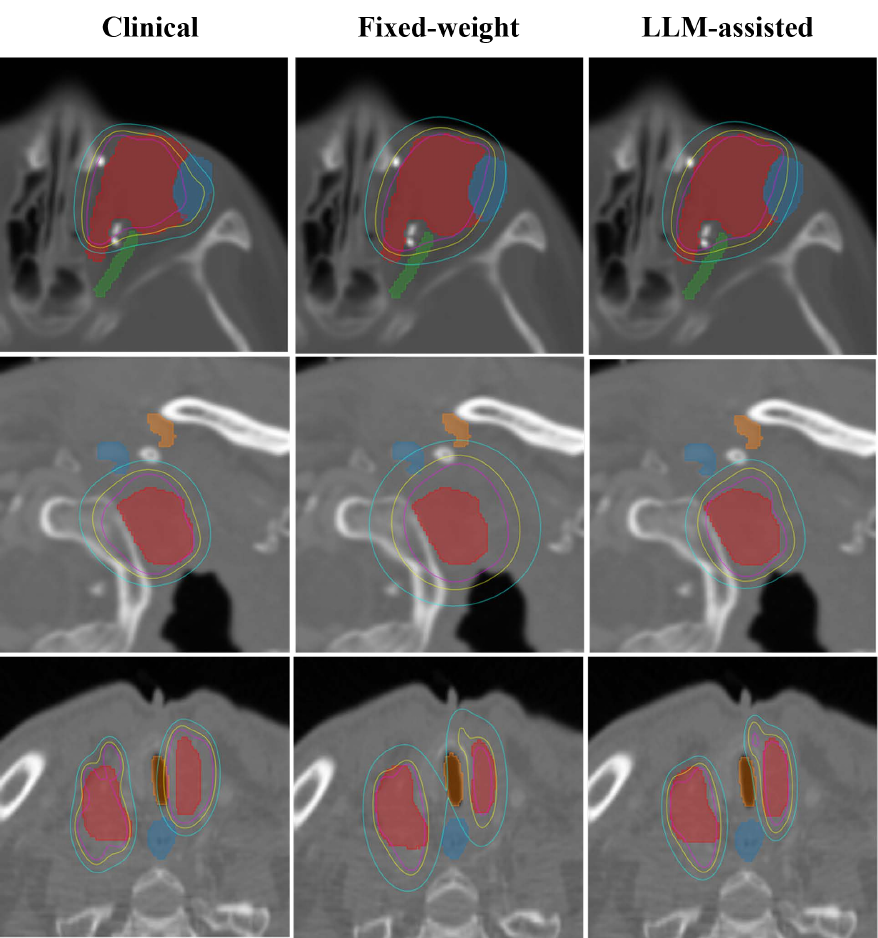} 
    \caption{Comparison of iso-dose distributions at 130\% (magenta), 100\% (yellow), and 70\% (cyan) of the prescription dose obtained by clinical plans (left column), fixed-weight plans (middle column), and LLM-assisted plans (right column). Three representative slices from different testing patient cases are shown. The red regions represent the CTVs. OARs are delineated with various colors, including the blue and orange regions.} 
    \label{fig:fig5} 
\end{figure}

Fig.~\ref{fig:fig5} compares iso-dose distributions for three representative patient cases across clinical, fixed-weight, and LLM-assisted adaptive plans. Fixed-weight plans show significant high-dose spillover, with expanded iso-dose regions extending beyond target boundaries, suggesting potential overexposure of adjacent tissues. Conversely, LLM-assisted adaptive plans demonstrate superior high-dose conformity to the CTVs, effectively limiting excessive dose delivery while ensuring adequate target coverage. Additionally, adaptive plans improve sparing of surrounding OARs, such as the trachea and esophagus in the third case, underscoring the clinical advantages of the proposed weight optimization framework.

\subsection{Ablation Study}

\begin{table*}[htbp]
\centering
% \captionsetup{justification=centering}
\caption{Ablation study of CTV dosimetric parameters, OAR sparing, and resource utilization with and without RAG.}

\begin{tabularx}{\textwidth}{l *{5}{>{\centering\arraybackslash}X}}
\toprule
Plans & $V_{100}$ (\%)↑ & $V_{150}$ (\%)↓ & $V_{200}$ (\%)↓ & $D_{90}$ (\%)↑ & \#needle ↓ \\
\midrule
w/o RAG & $95.4 \pm 3.4$ & $75.1 \pm 8.7$ & $54.0 \pm 11.1$ & $119.4 \pm 14.3$ & $9 \pm 2$ \\
w RAG & \textbf{96.4 $\pm$ 1.6} & \textbf{75.2 $\pm$ 6.2} & \textbf{51.8 $\pm$ 8.6} & \textbf{120 $\pm$ 7.9} & \textbf{9 $\pm$ 2} \\
$p$ value & $0.69$ & $0.34$ & $0.15$ & $0.43$ & $0.25$ \\
\midrule
\end{tabularx}

\vspace{1ex}

\begin{tabularx}{\textwidth}{l *{6}{>{\centering\arraybackslash}X}}
\toprule
Plans & $V_{50}$ (\%)↓ & $D_{\text{mean}}$ (Gy)↓ & $D_{\text{max}}$ (Gy)↓ & $D_1$ (Gy)↓ & $D_{1\text{cc}}$ (Gy)↓ & $D_{0.1\text{cc}}$ (Gy)↓ \\
\midrule
w/o RAG & $38.7 \pm 31.6$ & $47.4 \pm 24.9$ & $113.1 \pm 66.3$ & $102.9 \pm 50.5$ & $47.4 \pm 28.1$ & $82.5 \pm 40.1$ \\
w RAG & \textbf{36.3 $\pm$ 22.2} & \textbf{44.8 $\pm$ 16.5} & \textbf{109.5 $\pm$ 36.9} & \textbf{98.5 $\pm$ 29.8} & \textbf{47.0 $\pm$ 22.2} & \textbf{81.6 $\pm$ 25.3} \\
$p$ value & $0.31$ & $0.16$ & $0.22$ & $0.44$ & $0.16$ & $0.16$ \\
\bottomrule
\end{tabularx}

\vspace{0.5ex}
\begin{flushleft}
\textit{Note}: $P$ values are from two-sided Wilcoxon signed-rank tests comparing plans with RAG versus without RAG.
\end{flushleft}

\label{tab:ablation_rag}
\end{table*}

To further evaluate the contribution of the RAG module, we conducted an ablation study comparing plan quality with and without RAG integration. As summarized in Table~\ref{tab:ablation_rag}, while CTV coverage and needle number remained consistent between the two groups, the RAG-assisted plans demonstrated improved OAR sparing across multiple dosimetric metrics. These results suggest that incorporating retrieval-augmented clinical knowledge enhances the model’s ability to balance target coverage with normal tissue protection, contributing to more refined and clinically favorable treatment plans.

Without RAG, the average number of iterations until termination increased to 7.4 ± 3.1, with the selected plan located at iteration 5.3 ± 4.0. Notably, five cases reached the predefined upper limit of 10 iterations, indicating a higher likelihood of prolonged or suboptimal convergence when external clinical knowledge is not incorporated. 

\begin{figure*}[htbp]
    \centering
    \includegraphics[width=0.9\textwidth]{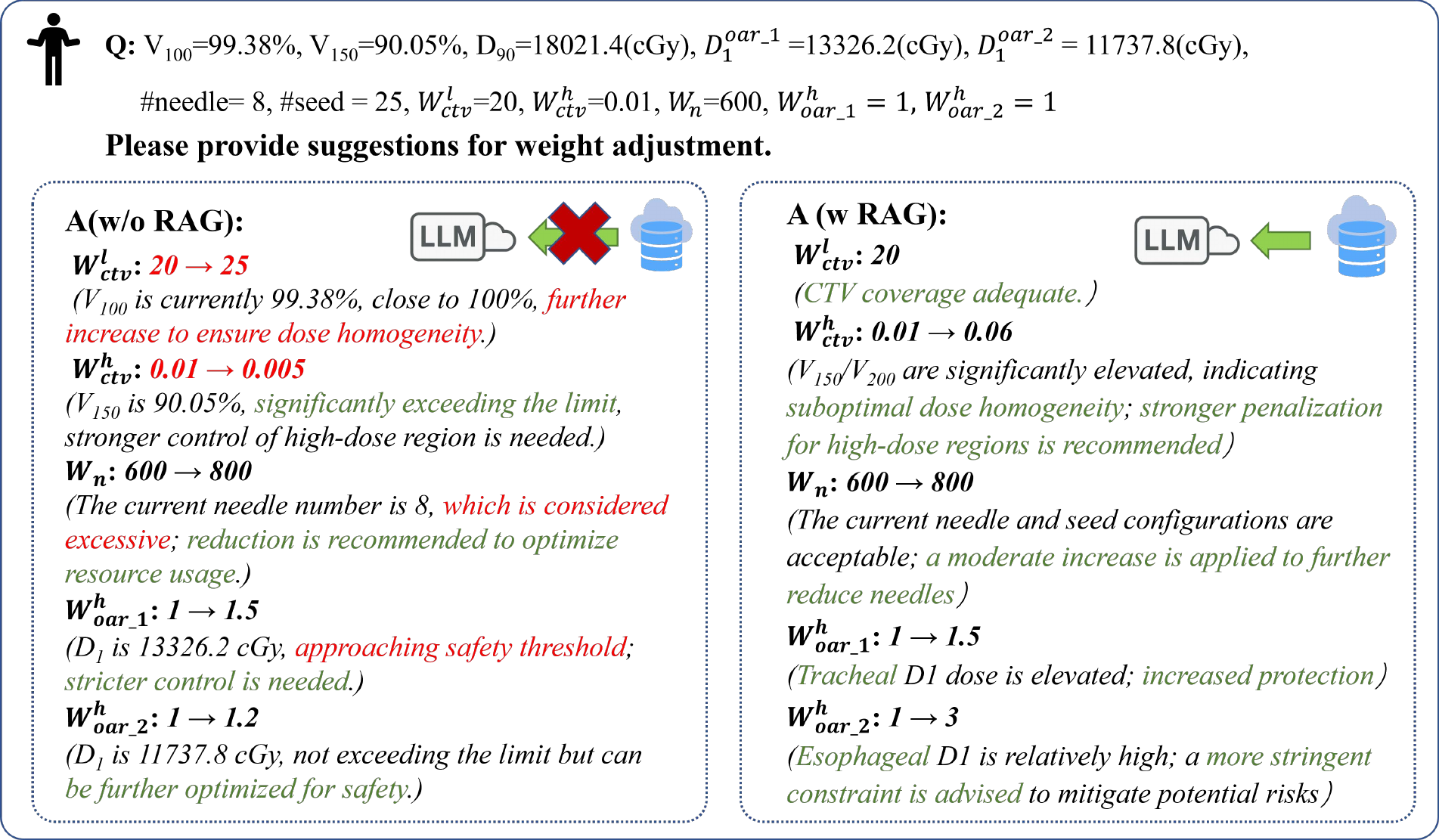} 
    \caption{Comparison of responses generated by LLM w/o (left) and w (right) RAG support. The hallucination-prone LLM without knowledge retrieval exhibits multiple clinically inconsistent suggestions. Correct and incorrect descriptions are highlighted in green and red, respectively.} 
    \label{fig:fig6} 
\end{figure*}

As illustrated in Fig.~\ref{fig:fig6}, we also compared the responses generated by the LLM during weight recommendation, with and without knowledge retrieval, to assess the impact of RAG on decision quality. The descriptions reflect the LLM's reasoning process and were manually verified for correctness.

\subsection{Computational efficiency}
In clinical practice, computational efficiency is essential to ensure timely treatment planning and support adaptive workflows. As previously described, the optimization process in our framework converged within 5.3 ± 0.8 iterations. All experiments were performed on a workstation equipped with an NVIDIA A100 GPU (80 GB VRAM) and an Intel Xeon w5-2465X CPU.

The total processing time was 3.7 ± 0.7 minutes per patient. Within each iteration, the heuristic solver required an average of 15.9 ± 13.4 seconds to converge, while the LLM-based evaluation and weight recommendation modules took 26.4 ± 8.4 seconds. Notably, the weight recommendation step involved processing both CTV and OAR DVH summary data as input, resulting in a larger token length and consequently longer inference time. In contrast, the evaluation step primarily relied on the optimization history, requiring fewer tokens and thus achieving relatively faster inference. Compared to conventional manual planning approaches, the proposed framework substantially reduces the overall planning time and operator workload.

\section{Discussion and conclusion}
This work presents a novel LLM-guided framework for adaptive weight optimization in SIBT planning. By integrating a locally deployed LLM model with an automated optimization engine in a closed-loop architecture, the system enables dynamic adjustment of objective weights based on plan evaluation feedback. To enhance the domain-specific reasoning capabilities of the LLM, a clinical knowledge base is incorporated and queried through a RAG mechanism, allowing the model to align its recommendations with established treatment guidelines. To the best of our knowledge, this study represents the first application of a locally deployed LLM to brachytherapy plan optimization. The approach significantly reduces manual parameter tuning while improving the dosimetric quality of treatment plans, particularly in terms of CTV coverage and OAR sparing.

One advantage of the proposed framework lies in its ability to operate entirely on local computational resources, which ensures data privacy and facilitates integration into clinical environments without reliance on external servers. To evaluate the feasibility of such deployment in real-world settings, we measured the actual planning time under local execution. Across all tested cases, the complete planning process for a single patient was completed within 3.7 ± 0.7 minutes, which is well within the acceptable range for routine clinical workflows. These results demonstrate that the framework remains computationally efficient for practical deployment despite the use of large-scale LLM and RAG components. Furthermore, with the application of concurrent LLM inference and model acceleration techniques, the overall processing time can be reduced.

Nevertheless, several aspects of the current framework remain open for further exploration and optimization. While the LLM assists weight adjustment, the underlying optimization solver still relies on existing heuristic algorithms\cite{zhangInversePlanningSimulated2022}, which are prone to stochastic perturbations and may be trapped in suboptimal local minima. An emerging research direction involves integrating LLM reasoning capabilities directly into the optimization process to guide search trajectories, thereby improving convergence efficiency and solution quality. Recent studies of such hybrid optimization paradigms that combine LLMs and traditional solvers have shown potential to further enhance planning robustness \cite{yangLARGELANGUAGEMODELS}\cite{wangLargeLanguageModel2024}.

Beyond weight adjustment, multimodal large language models also hold promise in needle trajectory design and evaluation. Although our current approach\cite{xiaoAutomaticPlanningHead2024} has achieved automated multi-needle placement, it still falls short of fully meeting clinical requirements and preferences, with certain trajectories requiring manual refinement. By incorporating imaging data, clinical context, and historical planning knowledge, MLLMs may facilitate more comprehensive decision-making and further reduce dependence on human expertise in needle trajectory planning, which remains a highly variable and complex aspect of clinical practice.

Finally, while the proposed framework demonstrates strong performance under experimental conditions, further studies are needed to validate its generalizability across different anatomical sites, treatment modalities, and clinical protocols. In addition, ensuring safety, reproducibility, and regulatory compliance are essential for clinical translation, and human-in-the-loop strategies\cite{buddSurveyActiveLearning2021} may remain necessary to supervise automated recommendations in high-stakes medical applications.

\bibliographystyle{plain}
\bibliography{reference}
\end{document}